\documentclass[twocolumn]{aastex62}

\usepackage{rotating}
\usepackage{booktabs}
\usepackage{dcolumn}
\usepackage{ulem}
\usepackage{breakcites}

\received{August 29, 2019}
\accepted{by ApJL November 21, 2019}
\shorttitle{Planetary engulfment events in binary systems}
\shortauthors{Nagar et al.}
\begin{document}

\title{The chemical signatures of planetary engulfment events in binary systems}

\correspondingauthor{Tushar Nagar}
\email{tusharnagar64@gmail.com}

\author[0000-0002-2747-0497]{Tushar Nagar}
\affil{Monash Centre for Astrophysics, School of Physics and Astronomy, Monash University, VIC 3800, Australia}

\author[0000-0002-9760-6249]{Lorenzo Spina}
\affil{Monash Centre for Astrophysics, School of Physics and Astronomy, Monash University, VIC 3800, Australia}

\author[0000-0002-3625-6951]{Amanda I. Karakas}
\affil{Monash Centre for Astrophysics, School of Physics and Astronomy, Monash University, VIC 3800, Australia}



\begin{abstract}

Planetary engulfment events involve the chemical assimilation of a planet into a star's external layer. This can cause a change in the chemical pattern of the stellar atmosphere in a way that mirrors the composition of the rocky object engulfed, with the refractory elements being more abundant than the volatiles. Due to these stellar chemical changes, planetary engulfment events can render the process of chemical tagging potentially inaccurate. A line-by-line differential analysis of twin stars in wide binary systems allows us to test the chemical homogeneity of these associations with typical individual stellar Fe~I uncertainties of 0.01 dex and eventually unveil chemical anomalies that could be attributed to planetary engulfment events. Out of the 14 systems analysed here, we report the discovery of the most chemically inhomogeneous system to date (HIP34407/HIP34426). The median difference in abundances of refractory elements within the pair is 0.19 dex and the trend between the differential abundances and condensation temperature suggests that the anomaly is likely due to a planetary engulfment event. Within our sample, five other chemically anomalous systems are found.

\end{abstract}




\section{Introduction} \label{sec:intro}
Within the last decade, radial velocity and transit surveys have discovered thousands of exoplanets around Sun-like stars. The global picture that has emerged exhibits a remarkable degree of diversity in terms of the architectures of these planetary systems \citep{Winn15}. Presumably, this observed diversity has arisen as a result of dynamical processes acting since the first stage of planetesimal formation \citep{Raymond18,Chambers18}. The fact that some systems have undergone complex phases of dynamical evolution is also attested to by the presence of planets on highly eccentric orbits (e.g., \citealt{Kane12}) that are misaligned or even counter rotating with respect to the spin axes of their hosting star (e.g., \citealt{Naoz11}), interstellar exoplanets (e.g., \citealt{Mroz18}) or by the observation of dusty debris disks formed through exoplanet collisions (e.g., \citealt{Kenyon16}). It is likely that in systems with evidence of a dynamical past, part of the planetary material has fallen into the hosting star (e.g., \citealt{Martinez19}), polluting its atmosphere and producing a significant increase in the stellar metallicity, which can be reliably detected (e.g., \citealt{Spina15,Spina18,Oh18,TucciMaia19}). In fact, such a dilution will not yield an indiscriminate abundance rise of all the heavy elements, but instead will produce a characteristic chemical pattern that mirrors the composition observed in rocky materials \citep{Chambers10, YanaGalarza16a, Kunitomo18} with most refractory elements (e.g., those having condensation temperature T$_{\rm cond}>$1000 K) being over-abundant relative to the volatiles (T$_{\rm cond}<$1000 K).

Prior proposals to explain these signatures suggest that the anomalous volatile to refractory ratios may be due to the selective accretion of volatiles. This could occur after the formation of rocky planets around a star \citep{Melendez09}. However, as already discussed in \citet{Spina18}, this explanation seems unlikely, as the chemical signature would have been imprinted on the star when it was too young (age$\leq$10 Myrs), diluting any accreted material due to the star's thick external layer.

Unveiling the chemical signatures of planetary engulfment events in stellar atmospheres is key to studying the frequency over which these catastrophic events occur, but it is also relevant to test the chemical homogeneity of stellar associations and probe the success of ``chemical tagging''. The concept of chemical tagging is to use the chemical information of stars to assign them to their progenitor cloud \citep{Freeman02}. A critical assumption of this technique is that members of the same stellar associations, such as open clusters or binary systems, are chemically identical as they formed at the same time and from the same material. Therefore, if the abundance of any element is altered due to a planetary engulfment event, the results from chemical tagging will no longer be reflective of the star's progenitor cloud.

In this letter we examine the chemical homogeneity of 14 wide binary star systems, making use of high-precision abundance determinations. In Section~\ref{sec:analysis} we describe the target selection, observations and method of analysis. The results of our study are presented in Section~\ref{sec:results} and our conclusions in Section~\ref{sec:conclusions}.

\section{Observations and analysis} \label{sec:analysis}

The 28 target stars were selected from lists of wide binary systems (separation $>$ 4'') compiled by \citet{Martin02}, \citet{Desidera04}, and \citet{Fuhrmann15}. Proper motions, radial velocities and parallaxes from Gaia DR2 \citep{GaiaDR2} are listed in Table~\ref{targets} and confirm that the pairs are physical systems. For this project, we have chosen only the pairs formed by twin dwarf stars with spectral types similar to that of the Sun. Twin stars are objects with atmospheric parameters very similar to each other (e.g., $\Delta$T$_{\rm eff}\lesssim$300~K and $\Delta$log~g$\lesssim$0.2 dex). It has been shown that a strictly differential line-by-line analysis of twin stars permits us to obtain differential abundances at the highest precision possible (e.g., \citealt{Melendez09,Bedell14,Bedell18,Liu2014, Teske2016, Spina18}). In fact, when studying samples of twin stars, most of the systematic uncertainties that plague chemical abundance analyses are so similar among stars in the same binary pair that a strict differential line-by-line analysis cancels them out. Also NLTE effects have been found negligible for all the elements considered by our analysis of solar twin stars \citep{Melendez12,Melendez2016,Nissen15,Spina16}, and so no NLTE corrections were required. This leaves the observational noise as the main source of error. Thus, error bars can be made very small simply by acquiring very high-quality spectra (i.e. a resolving power R$\geq$60,000 and signal-to-noise ratio S/N$\geq$300). 

Eight binary systems were observed by the Ultraviolet and Visual Echelle Spectrograph (UVES; \citealt{Dekker00}) on the Very Large Telescope of the European Southern Observatory\footnote{Data was used from ESO programme ID 0100.C-0090.}. The observations have been performed with a resolving power R$\sim$85,000 and a wavelength coverage between 330 - 680 nm, though the DIC-1 (390+580). Another six pairs have been observed with the High Dispersion Spectrograph (HDS; \citealt{Noguchi02}) on the Subaru telescope\footnote{Data was used from proposal ID o18123.}. For these latter we have used a R$\sim$80,000 and the Yc standard setup which covers the 439-705 nm spectral range. 
Thanks to the brightness of our targets (V$\leq$9.3 mag), we achieved a SNR ranging from 300-400 pixel$^{−1}$ at 600 nm, with a median of 350 pixel$^{−1}$. Solar spectra have been acquired both with UVES (S/N$\sim$350) and HDS (S/N$\sim$400) using the same instrument configurations described above.

All spectra have been normalized and Doppler-shifted using IRAF's \texttt{continuum} and \texttt{dopcor} tasks. Equivalent widths (EWs) of the atomic and molecular transitions reported in \citet{Melendez14} and listed in Table 2\footnote{Table 2 is available online at the CDS in its entirety.} have been measured with \texttt{Stellar diff}\footnote{\texttt{Stellar diff} is Python code publicly available at \url{https://github.com/andycasey/stellardiff}.}. This code allows the user to select one or more spectral windows for the continuum setting around each line of interest. Ideally, these windows coincide with regions devoid of other absorption lines. We employed the same window settings to calculate continuum levels and fit the lines of interest with Gaussian profiles in every spectrum. Therefore, the same assumptions have to be taken in the choice of the local continuum around the lines of interest. This is expected to minimize the effects of an imperfect spectral normalization or unresolved features in the continuum that can lead to larger errors in the differential abundances \citep{Bedell14}. Furthermore, \texttt{Stellar diff} is able to identify points affected by hot-pixels or cosmic rays and remove them from the calculation of the continuum. 
The code delivers the EW of each line of interest along with its uncertainty. 

The iron EW measurements are processed by the qoyllur-quipu (q2) code \citep{Ramirez14} that performs a line-by-line differential analysis relative to the Solar spectrum and automatically estimates the stellar parameters (effective temperature T$_{\rm eff}$, surface gravity log g, metallicity [Fe/H], and microturbulence $\xi$) by iteratively searching for the three equilibria: excitation, ionisation, and the trend between the iron abundances and the reduced equivalent width. We assumed the nominal solar parameters, T$_{\rm eff}$=5777 K, log g=4.44 dex, [Fe/H]=0.00 dex and $\xi$ =1.00 km s$^{-1}$ \citep{Cox00}. The iterations are executed with a series of steps starting from a set of initial parameters and employing the Kurucz (ATLAS9) grid of model atmospheres \citep{Castelli04}. In each step the abundances are estimated using MOOG (version 2014, \citealt{Sneden73}). The errors associated with the stellar parameters are then evaluated by the code. This takes into account the dependence between the parameters in the fulfillment of the three equilibrium conditions \citep{Epstein10}. We first run q2 adopting the Solar parameters as a first guess for each star. After q2 has converged to a set of stellar parameters, the differential abundances relative to the Sun for the following elements are calculated: C~I, CH, NH, Na~I, Mg~I, Al~I, Si~I, S~I, Ca~I, Sc~II, Ti~I, Ti~II, V~I, Cr~I, Cr~II, Mn~I, Fe~I, Fe~II, Co~I, Ni~I, Cu~I, Zn~I, Y~II, Zr~II, and Ba~II. Through the blends driver in MOOG and adopting the line list from the Kurucz database, the q2 code corrected the abundances of V, Mn, Co, Cu, and Y for hyperfine splitting effects, by using the HFS components in the input line list. For each element, we performed a 3-sigma clipping on the abundances yielded by each EW measurement. This allowed us to remove the EW measurements affected by telluric lines or other unresolved blendings with adjacent lines. A second run of q2 using the restricted list of EW measurements yielded the stellar parameters listed in Table~\ref{parameters}. Atmospheric T$_{\rm eff}$ values range from 5515 - 6295 K, while log~g values range within 3.975 - 4.523 dex. All pairs are composed of twin stars.

With these final parameters, we repeated the calculation of the differential abundances relative to the Sun. The resulting abundances are listed in Table 4, together with their uncertainties and the number of lines used for the abundance determinations. The error budget associated with each elemental abundance has been obtained by summing in quadrature the standard error of the mean among the lines, and the propagated effects of the uncertainties on the stellar parameters. The typical precision that we achieved in individual stellar Fe~I abundances is 0.01 dex. We also determined the differential abundances within each pair which are listed in Table 5\footnote{Tables 4 and 5 are available online at the CDS in their entirety.}.

The forbidden Oxygen line at 6300.3~$\AA$ has been used to derive [O/H] for eight stars in our sample. The other stars either had too small of an [OI] line sunk in the spectral noise or were contaminated by O$_{2}$ telluric lines. The measurement of this line requires particular care as it is small (typically 3 - 6~m$\AA$) and it is blended by a Ni line with nearly the same wavelength \citep{Allende-Prieto01}. Thus, we have followed a procedure already tested by \citet{Nissen15}. Namely, i) using Iraf's \texttt{Splot} task we measured the EW of [OI]+Ni line and its uncertainty by assuming different local continuum levels ; ii) through MOOG's \texttt{ewfind} task and [Ni/H] abundances in Table 4, we have calculated the EW contribution of the Ni line and its uncertainty; iii) we have subtracted the Ni contribution from the measured EW in order to estimate the EW of the [OI] line along with its uncertainty; iv) using the [OI] EW and the parameters listed in Table~\ref{parameters}, we calculated the [O/H] and its uncertainty. The Oxygen differential abundances are also listed in Tables 4 and 5. 

The differential abundances within each pair are plotted in Fig.~\ref{plots} as a function of the condensation temperature.

\section{Results} \label{sec:results}

In Table~\ref{Homogeneity_test} we report the reduced chi-square value $\chi^{2}_{red}$ of the pairs, which is defined as 
\begin{equation}
  \chi^{2}_{red} = \frac{1}{N-1}\sum_{i=1}^N  \left( \frac{\Delta[X_i/H]}{\sigma_{X_i}}  \right)^2,
\end{equation}
where $\Delta$[X$_i$/H] is the differential abundance of the X$_{i}^{th}$-element within the pair, $\sigma_{X_{i}}$ its uncertainty and N is the number of elements detected in the components of the pair. In order to assess if the pair is chemically anomalous or not, we can compare the $\chi^{2}_{red}$ values with the value expected for the model where the pair is assumed to be chemically identical: $\chi^{2}_{red}$=1. However, as noted by \citep{Andrae}, this value has an uncertainty $\sigma_\chi$=2/N which is given by the width of the $\chi$-distribution due to the random noise of the data (which we can assume to be Gaussian). The $\sigma_\chi$ values are also listed in Table~\ref{Homogeneity_test}. Consequently, the six pairs of our sample with $\chi^{2}_{red}>$1+3$\sigma_\chi$ are all considered chemically anomalous.

The HIP34407/HIP34426 system has a $\chi^{2}_{red}$ value of 118.99, indicating that the two stars differ considerably in chemical composition. Five other systems are chemically anomalous, although to a lesser extent. They are HIP44858/HIP44862, HD98744/HD98745, HD103431/HD103432, HD105421/HD105422, and \\ HIP70386A/HIP70386B, which have $\chi^{2}_{red}$ values equal to 2.47, 5.56, 4.92, 5.59, and, 1.95 respectively. Therefore, out of the 14 pairs considered in this letter, six could render chemical tagging potentially inaccurate at our level of precision. However, assuming uncertainties of 0.05 dex for each element, which is a typical value for large spectroscopic surveys (e.g., \citealt{Smiljanic14,Buder2018}), we calculated the $\chi^{2}_{\rm Survey}$, also listed in Table~\ref{Homogeneity_test}. Based on this, only two pairs out of the 14 would result in being chemically inhomogeneous if targeted by a spectroscopic survey: HIP34407/HIP34426 and HD98744/HD98745.

\begin{figure*}[ht]
\centering
\includegraphics[width=13cm]{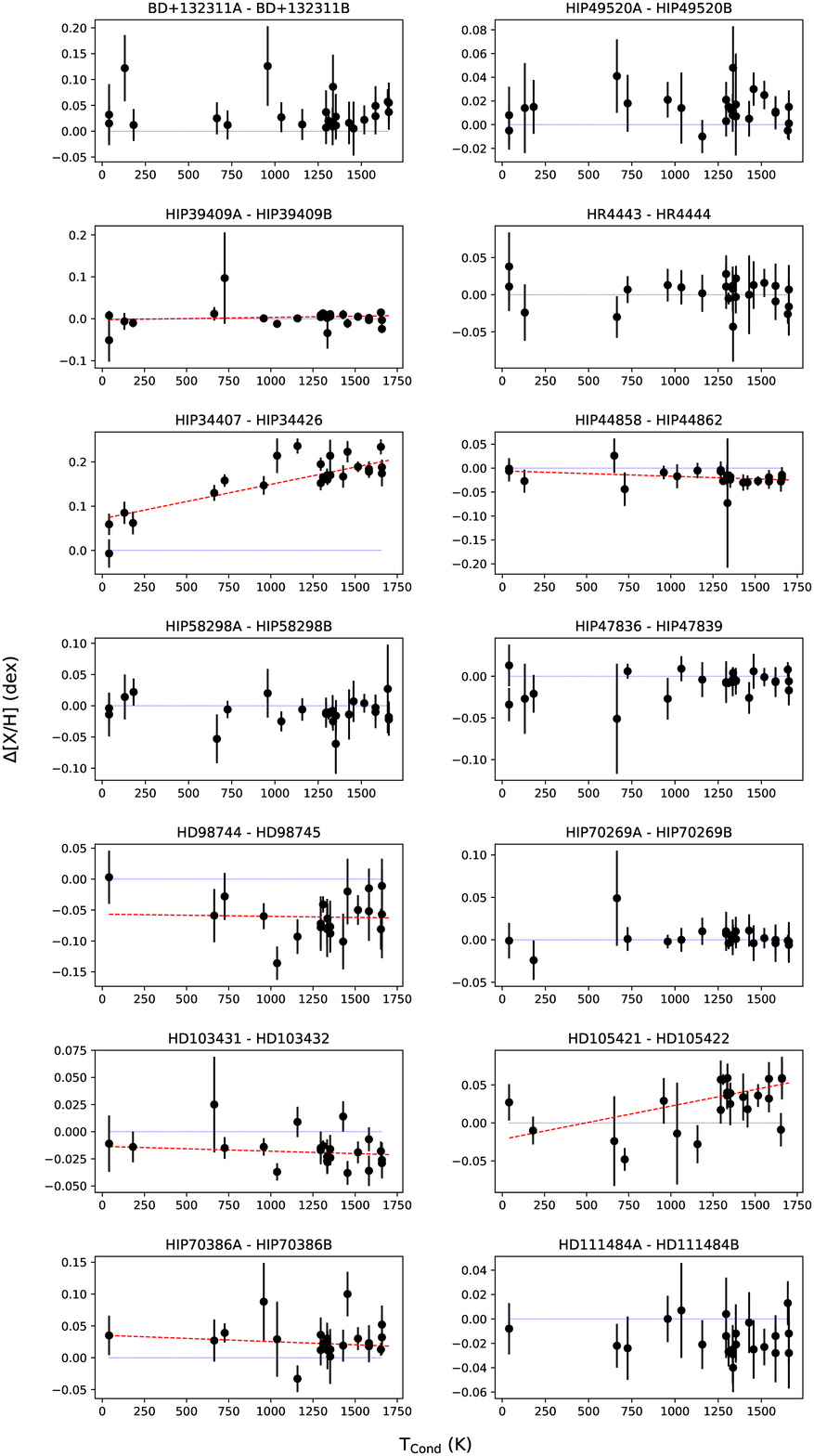}
\caption{Each panel shows the differential abundances for a single system as a function of the condensation temperature. The red dashed lines are the results of the linear fitting of the $\Delta$[X/H]-T$_{\rm Cond}$ distributions for the chemically anomalous pairs.
\label{plots}}
\end{figure*}

A rocky planet falling into a star and polluting its atmosphere would result in a selective enhancement of refractory elements \citep{Chambers01}. Therefore, in order to verify if the chemical anomalies identified above could be attributed to planetary engulfment events, we have performed a linear fit of the differential abundances as a function of the condensation temperature. The resulting slopes are also listed in Table~\ref{Homogeneity_test}. Among the initial six chemically anomalous pairs, three have a slope that is consistent with being zero, while another three pairs (i.e. HIP44858/HIP44862, HD105421/HD105422, and HIP34407/HIP34426) have a slope that is inconsistent with zero. However, it should be noted that most of the slopes listed in Table~\ref{Homogeneity_test} are heavily driven by the differential abundances of the most volatile species, such as C, CH, O, and N (i.e. T$_{\rm Cond}<$200~K). In fact, the number of volatiles is much smaller than the number of refractory elements, as is visible in Fig.~\ref{plots}. In addition, the chemical abundances of volatile elements are often more uncertain than those of refractories.

In the case of HIP34407/HIP34426, the chemical anomaly is extremely large compared to all the other pairs. The elements with an intermediate condensation temperature, such as Na, S, and Zn (i.e. 500$<$T$_{\rm Cond}<$1000~K) are well aligned by the $\Delta$[X/H]-T$_{\rm Cond}$ relation plotted in Figure.~\ref{plots}. This neat linear relation between  elemental abundances and condensation temperature for this system may suggest that the anomaly was caused by a planetary engulfment event. Similar results were independently obtained by \citet{Ramirez19}. 

Interestingly, the HIP34407/HIP34426 system was determined to be the most chemically inhomogeneous system found to date among other pairs in binary systems or clusters, that show similar trends between abundances and condensation temperature \citep{Spina15,Spina18,TucciMaia14,TucciMaia19,Teske15,Teske16,Oh18,Biazzo15,Saffe16,Saffe17}. Its median differential abundance in the refractory elements is 0.19 dex. In addition, we note that the results from the HIP34407/HIP34426 system involve similar but not identical volatile elemental abundances. Out of the four most volatile species measured (i.e. T$_{\rm Cond}<$200 K), only one is consistent with zero, leading to the hypothesis that the slope is due to the engulfment of a gas giant. This is because such planets are capable of causing abundance changes in both volatiles (due to their gaseous outer layers) and refractories (due to their rocky cores and metallic inner layers). This hypothesis has been also proposed to explain a similar anomaly observed in the chemical composition of the 16 Cyg binary system \citep{TucciMaia14, TucciMaia19}. 

\section{Conclusions} \label{sec:conclusions}

The occurrence of chemically anomalous stars at the 0.01 dex precision level can be deduced from line-by-line differential analysis of twin stars in binary systems. Out of the 14 systems measured using UVES and HDS, six were found to exhibit chemical inhomogeneities (i.e. $\chi^{2}_{red}>$1+3$\sigma_\chi$), rendering chemical tagging on these systems inaccurate. However, only two pairs were chemically inhomogeneous at the precision level typical of large spectroscopic surveys (i.e. 0.05 dex). 

Of special note is that the HIP34407/HIP34426 system was determined to be the most chemically inhomogeneous system found to date. For this pair, the trend between differential abundances and condensation temperature is neat, and it may suggest that rocky material has polluted the atmosphere of HIP34407. We speculated that the planet engulfed by the star was a giant gaseous planet, as HIP34426 is also anomalously rich in the volatiles. The planet population around the stars of this pair is currently unknown. Follow-up observations are required to establish if the two stars also have two different architectures for their planetary systems. This is necessary to understand the origin of similar anomalies observed within other stellar associations \citep{Spina15,Spina18,TucciMaia14,TucciMaia19,Teske15,Teske16,Oh18,Biazzo15,Saffe16,Saffe17}, where the stars richer in refractory elements could also be the ones with a more chaotic architecture of their planetary systems.
 
Finally, systems whose components have identical stellar parameters and chemical patterns were also observed, such as the HIP58298A/HIP58298B and HR4443/HR4444 systems. These systems are excellent laboratories to test if other quantities, such as the thickness of the convective zone, stellar rotation, and activity can differ (and to what extent) among stars with equal mass, age and chemical composition.

\vspace{3mm}

We thank the many scientists and engineers who
made the UVES and HDS observations possible. It is a pleasure to acknowledge M. Asplund, A.R. Casey, J. Mel\'{e}ndez and D. Yong for helpful discussions. L.~S. and A.~I.~K.  acknowledge financial support from the Australian Research Council (Discovery Project 170100521).

\vspace{3mm}

\facilities{VLT(UVES), Subaru(HDS).}


\software{ \texttt{qoyllur-quipu} \citep{Ramirez14}, \texttt{MOOG} \citep{Sneden73}, \texttt{Stellar diff} (\url{https://github.com/andycasey/stellardiff}).}

\setcounter{table}{0}
\begin{table*}[ht]
\centering
\caption{Target selection}
\label{targets}
\medskip
\begin{tabular}{ccccccc}
\hline
Star & \multicolumn{4}{c}{Gaia DR2 Kinematics} & SNR & Instrument \\
& Parallax & pmRA & pmDEC & Radial velocity & &  \\
& [mas] & [mas yr$^{-1}$] & [mas yr$^{-1}$] & [km s$^{-1}$] & &\\
\hline

Sun                            &                                                                                                               &                  &                  &                 &                               &                                      \\
BD+132311A                     & 6.07$\pm$0.05                                                                                                 & -56.44$\pm$0.08  & -41.78$\pm$0.06  & 26.33$\pm$1.07  & 400                           & UVES                                 \\
BD+132311B                     & 5.72$\pm$0.05                                                                                                 & -53.16$\pm$0.08  & -41.86$\pm$0.06  & 26.08$\pm$0.66  & 300                           & UVES                                 \\
HIP34407                       & 20.97$\pm$0.05                                                                                                & -51.63$\pm$0.09  & -206.35$\pm$0.08 & -12.64$\pm$0.19 & 300                           & UVES                                 \\
HIP34426                       & 20.87$\pm$0.05                                                                                                & -54.18$\pm$0.09  & -213.07$\pm$0.08 & -11.83$\pm$0.20 & 350                           & UVES                                 \\
HIP39409A                      & 15.39$\pm$0.04                                                                                                & -15.41$\pm$0.05  & -14.81$\pm$0.04  & 47.78$\pm$0.15  & 300                           & UVES                                 \\
HIP39409B                      & 15.31$\pm$0.04                                                                                                & -15.13$\pm$0.06  & -13.07$\pm$0.04  & ---             & 400                           & UVES                                 \\
HIP44858                       & 20.45$\pm$0.10                                                                                                & -53.24$\pm$0.13  & 71.66$\pm$0.10   & 30.02$\pm$0.22  & 300                           & UVES                                 \\
HIP44864                       & 20.36$\pm$0.09                                                                                                & -51.82$\pm$0.11  & 73.52$\pm$0.07   & 30.31$\pm$0.14  & 350                           & UVES                                 \\
HIP47836                       & 17.37$\pm$0.04                                                                                                & -23.58$\pm$0.07  & 97.16$\pm$0.07   & -2.25$\pm$0.22  & 300                           & UVES                                 \\
HIP47839                       & 17.49$\pm$0.04                                                                                                & -21.34$\pm$0.07  & 98.41$\pm$0.06   & -1.94$\pm$0.21  & 300                           & UVES                                 \\
HIP49520A                      & 17.07$\pm$0.05                                                                                                & 41.09$\pm$0.07   & -57.17$\pm$0.07  & -0.56$\pm$0.67  & 300                           & UVES                                 \\
HIP49520B                      & 17.05$\pm$0.12                                                                                                & 56.91$\pm$0.17   & -52.46$\pm$0.17  & 1.11$\pm$0.45   & 350                           & UVES                                 \\
HIP58298A                      & 13.16$\pm$ 0.06                                                                                               & 35.76$\pm$ 0.09  & -65.41$\pm$ 0.04 & -3.62$\pm$ 0.27 & 300                           & UVES                                 \\
HIP58298B                      & 13.06$\pm$ 0.06                                                                                               & 37.08$\pm$0.10   & -71.90$\pm$ 0.04 & -3.35$\pm$ 0.29 & 300                           & UVES                                 \\
HR4443                         & 35.98$\pm$0.13                                                                                                & -22.03$\pm$0.19  & 139.98$\pm$0.15  & 7.15$\pm$0.53   & 350                           & UVES                                 \\
HR4444                         & 36.06$\pm$0.13                                                                                                & -19.60$\pm$0.19  & 144.52$\pm$0.16  & ---             & 350                           & UVES                                 \\
HD98744                        & 5.55$\pm$0.06                                                                                                 & 35.39$\pm$0.10   & -36.02$\pm$0.17  & ---             & 350                           & HDS                                  \\
HD98745                        & 4.57$\pm$0.31                                                                                                 & 37.54$\pm$2.15   & -36.15$\pm$2.19  & ---             & 350                           & HDS                                  \\
HD103431                       & 25.25$\pm$0.05                                                                                                & -450.60$\pm$0.08 & -15.50$\pm$0.06  & 5.76$\pm$0.22   & 400                           & HDS                                  \\
HD103432                       & 25.24$\pm$0.05                                                                                                & -450.50$\pm$0.09 & -16.55$\pm$0.07  & 5.94$\pm$0.29   & 400                           & HDS                                  \\
HD105421                       & 18.77$\pm$0.04                                                                                                & -177.94$\pm$0.05 & -20.80$\pm$0.07  & 7.64$\pm$0.16   & 400                           & HDS                                  \\
HD105422                       & 18.94$\pm$0.05                                                                                                & -182.04$\pm$0.06 & -28.01$\pm$0.06  & ---             & 400                           & HDS                                  \\
HD111484A                      & 11.45$\pm$0.05                                                                                                & -76.70$\pm$0.10  & -3.73$\pm$0.06   & -21.10$\pm$0.20 & 400                           & HDS                                  \\
HD111484B                      & 11.45$\pm$0.44                                                                                                & -79.88$\pm$0.08  & -4.47$\pm$0.06   & -19.34$\pm$0.31 & 300                           & HDS                                  \\
HIP70269A                      & 23.86$\pm$0.04                                                                                                & 5.97$\pm$0.08    & -136.66$\pm$0.07 & -32.39$\pm$0.16 & 400                           & HDS                                  \\
HIP70269B                      & 23.77$\pm$0.04                                                                                                & 10.88$\pm$0.07   & -133.74$\pm$0.06 & -33.06$\pm$0.16 & 400                           & HDS                                  \\
HIP70386A                      & 26.29$\pm$0.10                                                                                                & 65.59$\pm$0.15   & -0.868$\pm$0.11  & -0.05$\pm$0.92  & 400                           & HDS                                  \\
HIP70386B                      & 26.54$\pm$0.10                                                                                                & 66.19$\pm$0.13   & 0.60$\pm$0.12    & ---             & 350                           & HDS                                  \\ 
\hline
\end{tabular}
\end{table*}

\setcounter{table}{1}
\begin{table*}[ht]
\centering
\caption{Equivalent width measurements - full table available online at the CDS}
\medskip
\begin{tabular}{ccccccc}
\hline
Wavelength & Species & E.P. & Log gf & Sun (UVES) & BD+132311A & ... \\
$[$\AA$]$ & & [eV] & & [m$\AA$] & [m$\AA$] & ... \\
\hline
4365.896   & 26.0      & 2.990   & -2.250 & 51.6 & 28.6 & ... \\
4389.245   & 26.0      & 0.052  & -4.583 & 71.3 & 46.6 & ... \\
4445.471   & 26.0      & 0.087  & -5.441 & 38.9 & --- & ... \\
4950.106   & 26.0      & 3.417  & -1.560 & 76.2 & 50.8 & ...  \\
4994.129   & 26.0      & 0.915  & -3.080 & 104.5 & 83.3 & ...  \\
5044.211   & 26.0      & 2.851 & -2.058 & 72.1 & 47.9 & ... \\
5054.642   & 26.0      & 3.640   & -1.921 & 39.8 & 18.0 & ... \\
5127.359   & 26.0      & 0.915  & -3.307 & 97.4 & 76.3 & ... \\
5127.679   & 26.0      & 0.052  & -6.125 & 19.3 & --- & ... \\
... & ... & ... & ... & ... & ... & ... \\
\hline
\end{tabular}
\end{table*}

\setcounter{table}{2}
\begin{table*}[ht]
\centering
\caption{Stellar parameters}
\label{parameters}
\medskip
\begin{tabular}{ccccc}
\hline
Star & T$_{\rm eff}$ & log~g & [Fe/H] & $\xi$ \\
& [K] & [dex] & [dex] & [km s$^{-1}$] \\
\hline

%Sun        & 5777                                             & 4.44                                            & 0.00                                 & 1.00                            \\
BD+132311A & 6275$\pm$27                                      & 4.117$\pm$0.05                                 & -0.224$\pm$0.02                       & 1.67$\pm$0.05                   \\
BD+132311B & 6295$\pm$28                                      & 4.135$\pm$0.06                                 & -0.225$\pm$0.02                     & 1.54$\pm$0.06                   \\
HIP34407   & 5988$\pm$8                                      & 4.378$\pm$0.03                                 & -0.335$\pm$0.01                     & 1.36$\pm$0.02                   \\
HIP34426   & 6047$\pm$15                                      & 4.36$\pm$0.03                                  & -0.506$\pm$0.02                     & 1.57$\pm$0.04                  \\
HIP39409A  & 5620$\pm$5                                       & 4.512$\pm$0.02                                 & 0.044$\pm$0.01                      & 0.98$\pm$0.02                   \\
HIP39409B  & 5587$\pm$6                                       & 4.477$\pm$0.02                                 & 0.035$\pm$0.01                      & 0.95$\pm$0.02                   \\
HIP44858   & 5996$\pm$12                                      & 4.507$\pm$0.02                                 & -0.324$\pm$0.01                     & 1.22$\pm$0.02                   \\
HIP44862   & 5595$\pm$10                                      & 4.508$\pm$0.03                                 & -0.307$\pm$0.01                     & 1.21$\pm$0.02                   \\
HIP47836   & 6072$\pm$11                                      & 4.407$\pm$0.03                                 & -0.310$\pm$0.01                     & 1.42$\pm$0.03                   \\
HIP47839   & 6149$\pm$12                                      & 4.402$\pm$0.04                                 & -0.306$\pm$0.01                     & 1.5$\pm$0.04                    \\
HIP49520A  & 5915$\pm$10                                      & 4.523$\pm$0.02                                 & -0.182$\pm$0.01                     & 1.13$\pm$0.03                   \\
HIP49520B  & 5846$\pm$8                                      & 4.522$\pm$0.02                                 & -0.196$\pm$0.01                     & 1.06$\pm$0.03                   \\
HIP58298A  & 6171$\pm$18                                     & 4.292$\pm$0.04                                 & -0.46$\pm$0.01                      & 1.79$\pm$0.05                   \\
HIP58298B  & 6177$\pm$18                                      & 4.323$\pm$0.05                                  & -0.455$\pm$0.01                     & 1.82$\pm$0.05                   \\
HR4443     & 6216$\pm$21                                      & 4.21$\pm$0.05                                 & 0.017$\pm$0.01                      & 1.59$\pm$0.04                   \\
HR4444     & 6201$\pm$19                                      & 4.183$\pm$0.04                                 & -0.028$\pm$0.01                     & 1.61$\pm$0.03 \\
HD98744    & 6195$\pm$21                                      & 3.975$\pm$0.07                                & -0.309$\pm$0.01                     & 1.78$\pm$0.05                   \\
HD98745    & 6223$\pm$31                                      & 4.301$\pm$0.09                                 & -0.228$\pm$0.02                     & 1.62$\pm$0.04                   \\
HD103431   & 5515$\pm$6                                       & 4.445$\pm$0.03                                 & -0.158$\pm$0.01                     & 0.85$\pm$0.02                   \\
HD103432   & 5644$\pm$7                                       & 4.497$\pm$0.02                                 & -0.13$\pm$0.005                      & 0.89$\pm$0.01                   \\
HD105421   & 6265$\pm$13                                      & 4.49$\pm$ 0.03                                 & -0.096$\pm$0.01                     & 1.47$\pm$0.02                   \\
HD105422   & 6014$\pm$15                                      & 4.497$\pm$0.03                                 & -0.139$\pm$0.01                     & 1.19$\pm$0.02                   \\
HD111484A  & 6249$\pm$15                                      & 4.463$\pm$0.04                                 & 0.104$\pm$0.01                      & 1.42$\pm$0.02                   \\
HD111484B  & 6243$\pm$14                                      & 4.405$\pm$0.04                                 & 0.125$\pm$0.01                       & 1.4$\pm$0.02                    \\
HIP70269A  & 5968$\pm$11                                      & 4.385$\pm$0.03                                 & -0.23$\pm$0.01                      & 1.24$\pm$0.02                   \\
HIP70269B  & 5986$\pm$9                                      & 4.403$\pm$0.04                                 & -0.24$\pm$0.01                       & 1.26$\pm$0.03                   \\
HIP70386A  & 6117$\pm$18                                      & 4.507$\pm$0.04                                 & -0.043$\pm$0.01                     & 1.45$\pm$0.03                   \\
HIP70386B  & 5982$\pm$11                                      & 4.508$\pm$0.03                                 & -0.066$\pm$0.01                     & 1.21$\pm$0.02                   \\
\hline
\end{tabular}
\end{table*}

\setcounter{table}{3}
\begin{table*}[ht]
\centering
\caption{Differential abundances relative to the Sun - full table available online at the CDS. The number in brackets is the number of lines measured.}
\label{diffabusun}
\medskip
\begin{tabular}{ccccccc}
\hline
Star       & [CI/H]             & [CH/H]             & [NH/H]             & [OI/H]             & [NaI/H] & ... \\
& [dex] & [dex] & [dex] & [dex] & [dex] & ...\\
\hline
BD+132311A & -0.17$\pm$0.02 (3) & -0.23$\pm$0.04 (1) & -0.05$\pm$0.04 (1) & -0.022$\pm$0.008 (1)  & -0.25$\pm$0.02 (3) & ... \\
BD+132311B & -0.19$\pm$0.03 (3)  & -0.27$\pm$0.05 (1) & -0.18$\pm$0.05 (1)  & 0.01$\pm$0.013 (1)   & -0.37$\pm$0.09 (3) &  ...  \\
HIP34407   & -0.337$\pm$0.017 (3) & -0.468$\pm$0.009 (4) & -0.62$\pm$0.15 (2) & -0.040$\pm$0.007 (1)  & -0.356$\pm$0.019 (3) &  ...   \\
HIP34426   & -0.330$\pm$0.019 (3)  & -0.53$\pm$0.02 (4)  & -0.55$\pm$0.02 (1) & -0.102$\pm$0.016 (1) & -0.503$\pm$0.008 (3) &  ...  \\
HIP39409A  & -0.055$\pm$0.018 (3) & -0.023$\pm$0.006 (4) & -0.05$\pm$0.03 (2) & 0.017$\pm$0.006 (1)  & -0.012$\pm$0.003 (3) &  ...  \\
HIP39409B  & -0.02$\pm$0.06 (3) & -0.032$\pm$0.009 (4) & -0.046$\pm$0.015 (2) & 0.027$\pm$0.008 (1)  & -0.012$\pm$0.006 (3) &  ...   \\
HIP44858   & -0.337$\pm$0.017 (3) & -0.388$\pm$0.016 (4) & -0.43$\pm$0.11 (2) & ---             & -0.395$\pm$0.019 (3) &  ...   \\
HIP44862   & -0.32$\pm$0.03 (3) & -0.386$\pm$0.014 (4) & -0.41$\pm$0.13 (2) & ---             & -0.383$\pm$0.008 (3) &  ...  \\
HIP47836   & -0.235$\pm$0.013 (3) & -0.322$\pm$0.015 (4) & -0.32$\pm$0.08 (2)  & 0.000$\pm$0.012 (1)          & -0.302$\pm$0.007 (3) &  ...   \\
HIP47839   & -0.25$\pm$0.03 (3) & -0.289$\pm$0.017 (4) & -0.29$\pm$0.11 (2) & 0.021$\pm$0.014 (1)  & -0.28$\pm$0.02 (3)  &  ...   \\
HIP49520A  & -0.228$\pm$0.015 (3) & -0.235$\pm$0.016 (3) & -0.29$\pm$0.11 (2) & -0.079$\pm$0.0017 (1) & -0.266$\pm$0.015 (3) &  ... \\
... & ... & ... & ... & ... & ... & ... \\
\hline
\end{tabular}
\end{table*}

\setcounter{table}{4}
\begin{table*}[ht]
\centering
\caption{Differential abundances within each pair - full table available online at the CDS.  The number in brackets is the number of lines measured.}
\label{diffabupair}
\medskip
\begin{tabular}{cccccccc}
\hline
Star1       & Star 2 & [CI/H]             & [CH/H]             & [NH/H]             & [OI/H]             & [NaI/H] & ... \\
& & [dex] & [dex] & [dex] & [dex] & [dex] & ... \\
\hline
BD+132311A & BD+132311B & -0.02$\pm$0.03 (3) & -0.03$\pm$0.06 (1) & -0.12$\pm$0.06 (1) & -0.01$\pm$0.03 (1) & -0.13$\pm$0.08 (3)& ...  \\
HIP34407   & HIP34426   & 0.01$\pm$0.03 (3)  & -0.06$\pm$0.02 (4) & -0.09$\pm$0.03 (1) & -0.06$\pm$0.03 (1) & -0.15$\pm$0.02 (3) & ...  \\
HIP39409A  & HIP39409B  & 0.05$\pm$0.05 (3)  & -0.008$\pm$0.011 (4) & 0.01$\pm$0.02 (2) & 0.010$\pm$0.010 (1) & -0.001$\pm$0.005 (3) & ...  \\
HIP44858   & HIP44862   & 0.01$\pm$0.02 (3)  & 0.00$\pm$0.02 (4)  & 0.03$\pm$0.02 (2)   & ---                & 0.009$\pm$0.014 (3)  & ... \\
HIP47836   & HIP47839   & -0.01$\pm$0.03 (3) & 0.03$\pm$0.02 (4)  & 0.03$\pm$0.04 (2)  & 0.02$\pm$0.02 (1)  & 0.03$\pm$0.03 (3)& ...   \\
HIP49520A  & HIP49520B  & 0.005$\pm$0.016 (3)  & -0.01$\pm$0.02 (3) & -0.01$\pm$0.04 (2) & -0.02$\pm$0.02 (1) & -0.021$\pm$0.015 (3)& ...  \\
HIP58298A  & HIP58298B  & 0.00$\pm$0.02 (2)  & 0.01$\pm$0.04 (4)  & -0.01$\pm$0.04 (1) & -0.02$\pm$0.02 (1) & -0.02$\pm$0.04 (3)& ...  \\
HR4443     & HR4444     & -0.04$\pm$0.05 (3) & -0.01$\pm$0.03 (1) & 0.02$\pm$0.04 (1)  & ---                & -0.01$\pm$0.02 (2)& ...  \\
HD98744    & HD98745    & 0.00$\pm$0.04 (4)  & ---                & ---                & ---                & 0.06$\pm$0.02 (3)& ...  \\
... & ... & ... & ... & ... & ... & ... & ... \\

\hline

\hline
\end{tabular}
\end{table*}

\setcounter{table}{5}
\begin{table*}[ht]
\centering
\caption{Parameters for the chemical homogeneity test}
\label{Homogeneity_test}
\medskip
\begin{tabular}{ccccc}
\hline
Binary System & $\chi^{2}_{red}$ & $\chi^{2}_{Survey}$ & $\sigma_\chi$ & $\Delta$[X/H] vs T$_{\rm Cond}$ slope \\
& & & & 10$^{-6}$ dex K$^{-1}$\\ \hline
BD+132311A/BD+132311B & 1.10 & 0.93 & 0.28 & --- \\
HIP49520A/HIP49520B & 1.35 & 0.14 & 0.28 & --- \\
HIP39409A/HIP39409B & 2.07 & 0.25 & 0.28 & 5.6$\pm$5.9 \\
HR4443/HR4444 & 0.57 & 0.14 & 0.28 & --- \\
HIP34407/HIP34426 & 118.99 & 12.07 & 0.28 & 77.6$\pm$13.0 \\
HIP44858/HIP44862 & 2.47 & 0.28 & 0.28 & $-$11.2$\pm$5.4 \\
HIP58298A/HIP58298B & 0.66 & 0.19 & 0.28 & --- \\
HIP47836/HIP47839 & 0.62 & 0.12 & 0.28 & --- \\
HD98744/HD98745 & 5.56 & 1.99 & 0.30 & $-$3.6$\pm$23.0 \\
HIP70269A/HIP70269B & 0.19 & 0.06 & 0.29 & --- \\
HD103431/HD103432 & 4.92 & 0.21 & 0.29 & $-$4.5$\pm$8.0 \\
HD105421/HD105422 & 6.59 & 0.63 & 0.29 & 43.7$\pm$15.4 \\
HIP70386A/HIP70386B & 1.95 & 0.62 & 0.30 & $-$10.1$\pm$12.5 \\
HD111484A/HD111484B & 1.28 &  0.18 & 0.30 & --- \\
\hline
\end{tabular}
\end{table*}



%\bibliography{sample63}{}
%\bibliographystyle{aasjournal}

%\begin{thebibliography}{}
\clearpage
\bibliography{Bibliography} 
%\end{thebibliography}

\end{document}